\documentclass[twocolumn,prl,amsmath,amssymb,showpacs]{revtex4}
\usepackage{mathrsfs}
\usepackage{graphicx}% Include figure files
\usepackage{dcolumn}% Align table columns on decimal point
\usepackage{bm}% bold math
\usepackage{slashed}% \slashed{A}~a\!\!\!/

 {\catcode`\|=\active \gdef|{\egroup\,\vrule\,\bgroup}}

\def\be{\begin{eqnarray}}
\def\ee{\end{eqnarray}}
\begin{document}

%\preprint{PRL/Version 1}
\preprint{OL/Version 1}

%\title{Muon Decay in a Strong Laser Field}% Force line breaks with \\
\title{Laser-modified Angular Distribution of Muon Decay}% Force line breaks with \\
\author{Ai-Hua Liu}
% and Shu-Min Li}%
%
\affiliation{Department of Modern Physics,  University of Science
and Technology of China,  P.  O.  Box 4,  Hefei,  Anhui 230026,
People's Republic of China}

\author{Shu-Min Li}%
\affiliation{Department of Modern Physics,  University of Science
and Technology of China,  P.  O.  Box 4,  Hefei,  Anhui 230026,
  People's Republic of China and \\
China Center of Advanced Science and Technology (World
Laboratory),
 P. O. Box  8730,  Beijing  100080,  People's  Republic  of
 China}
%
%\author{Jamal Berakdar}%
%%\email{jber@mpi-halle.de}
%\affiliation{Institut f\"ur Physik, Martin-Luther-Universit\"at
%Halle-Wittenberg, Nanotechnikum-Weinberg, Heinrich-Damerow-Str.~4,
%06120 Halle
%}%
%
%

%\author{Charlie Author}
% \homepage{http://www.Second.institution.edu/~Charlie.Author}
%\affiliation{
%Second institution and/or address\\
% This line break forced% with \\
% }%

\date{\today}% It is always \today, today,
             %  but any date may be explicitly specified

\begin{abstract}
{We show theoretically that the angular distribution of decay rate
of muon can be changed dramatically by embedding the decaying muon
in a strong linearly polarized laser field. Evaluating the
S-matrix elements taking all electronic multiphoton processes into
account. The results suggest the muon may have internal structures
instead structureless as in the standard model. }
\end{abstract} \vskip 0.5in

%14.60.Cd, %Electrons (including positrons)
%14.60.Ef, %Muons
%13.40.Ks, % Electromagnetic corrections to strong- and weak-interaction processes
%13.35.Bv, % Decays of muons
%12.15.-y, %Electroweak interactions
%42.62.-b, %Laser applications
\vskip
1in
\maketitle
\emph{Introduction.}- Recent advances in the generation and
control of ultra-intense laser fields \cite{sj0} paved the way for
a number of spectacular applications ranging from  particles
accelerations \cite{sj3,sj1} and the generation of $X$-ray pulses
\cite{sj7} to laser-driven nuclear reaction \cite{sj8}, and
laboratory astrophysical and high-energy processes \cite{sj9}. In
addition to laser-induced phenomena, modifications of the
properties of elementary particles due to the presence of a strong
light field are enjoying much of attention. To name but one,
Chelkowski \emph{et al.} \cite{Chelkowski} demonstrated
theoretically that upon the ionization and dissociation of muonic
molecular ions in superintense laser fields (with intensities $
I\sim 10^{21}\: Wcm^{-2}$) the recolliding ions can ignite a
nuclear reaction with sub-laser-cycle precision and serve hence as
precursors for laser-assisted nuclear processes. A crucial
assumption in this kind of studies is that the participating
elementary particles are stable during the collision process. The
purpose of this work is to point out  that the particles
field-free life time may change dramatically  due to the presence
of a strong electromagnetic field (field amplitude $\sim 10^{6}\
Vcm^{-1}$). To accomplished such a process in experiment, we may
shed a laser beam on the beam of muon. The strong field may also
be used to affect the decay of other unstable particles, and
reveal new aspects of the decay mechanism.  This aspect of the
laser-matter interaction is of a prime importance, e.g. when
considering strong-field assisted collisions.

Specifically, we consider the modification of the decay life time
of muons due to the presence of a strong laser field. As well
known, muons have  played a crucial  role in the development and
assessment of the standard model \cite{Griffiths} and the
field-free muon decay was the first from all leptonic processes to
be investigated in full details \cite{Brom}.
%The current measured value of the muon lifetime is
%$\kern.05ex{\tau_{\mu}=(2.19703\pm0.00004)\times10^{-6} sec}$
%\cite{PDG06}. As established theoretically \cite{Greiner},
%radiative corrections (bremsstrahlung diagrams, the emission of an
%extremely ``soft" photon from a decay without radiation, etc.)
%contribute  to the decay rate by a factor
%$1-\frac{\alpha}{2\pi}\left(\pi^{2}-\frac{25}{4}\right)=0.9958\cdots$,
%where $\alpha=1/137$ is the fine structure constant. \textbf{The
%calculation on radiative decay of muon convince us that the
%calculation in this letter on the laser assistance on muon decay
%is credible and may be verified by experiment sooner.}

%(see Fig.\ref{fey:1}).
%
%\begin{figure}
%%[b]
%\includegraphics[height=6  cm]{fey1.eps}
%\caption{The Feynman diagram for the radiative correction in muon
%decay. } \label{fey:1}
%\end{figure}

The field-free muon lifetime may be modified by a number of
factors: E.g., Czarnecki \emph{et~al.} \cite{Czar00} investigated
the modifications of the $\mu^{+}$ lifetime  due to muonium
($\mu^{+}e^{-}$) formation and other medium effects, whereas
Vshivtsev and \'{E}minov \cite{Vshi79} studied the influence of  a
weak field on $\tau_\mu$.
In recent years, few investigations of the muon or muonium-laser
interactions were carried out:  Chu \emph{et~al.} studied  the
laser excitation of the muonium $1S-2S$ transition \cite{Chu88}
whereas Nagamine \emph{et~al.} \cite{Naga95} reported on an
ultraslow $\mu^{+}$ generation upon laser ionization of thermal
muonium. Our focus in this work is on the decay of the muon into
an electron, a muon-neutrino $\nu_{\mu}$,  and an electronic
antineutrino $\bar{\nu}_{e}$ in a strong laser field. Here we note
that present-day laser sources produce intensities of $10^{18}\,
Wcm^{-2}$ or higher in which case the averaged quiver energy of
the electron in the laser field may well exceed its rest energy
\cite{Milc96} necessitating  thus  a full relativistic treatment.

\emph{Theoretical formulation.}-
We assume the decay of a muon to occur in the presence of a
monochromatic, linearly polarized, spatially homogeneous laser
field. The final state electron is treated relativistically. The
electron energy ranges from $m_e$ to $m_{\mu}/2$ where $m_e$ and
$m_{\mu}$ are respectively  the rest masses  of the electron and
the muon. The laser is supposed to be switched on adiabatically
for a duration considerably longer than $\tau_\mu$. The laser
intensity is chosen such that  pair creation \cite{Bula96} is
negligible. The electromagnetic field is described by the
classical four-potential
 (unless otherwise stated, we use natural units  in which
%$\hbar=1=c$)
$\hbar=c=1$)
  ${A(x)=a\cos(k{\cdot} x)}$ that satisfies the Lorenz
condition.
%$\partial
%A(x) = 0$ is described by(linear polarization):
% \begin{equation}
%    {A(x)=a\cos(kx)} \label{eq:as},
%    \end{equation}
%The constant four vector $a = (0,{\bf a})$ stands for the vector potential
%${\bf a}=\mbox{\boldmath${\cal E}$}_0/\omega$, where
The constant four vector $a = (0, \mbox{\boldmath${\cal
E}$}_0/\omega)$, where $\mbox{\boldmath${\cal E}$}_0$ is the
amplitude of the electric field strength of laser. The wave four
vector $k = ({\omega}, {\bf k})$ follows from the laser frequency
$\omega$ and wave number ${\bf k}$.

The $S$-matrix elements for the laser-assisted $\mu^{-}$ decay
%$\mu^-\to e^- +\nu_\mu+\overline\nu_{e}$ process
%is
reads \cite{Brom,Greiner}
%    \begin{eqnarray}
%S_{fi}&=& -i\int_{-\infty}^{\infty}
%%\mathscr
%{H}_I(\mu^-\to e^-\nu_\mu\overline\nu_{e}) \; dt
%    \label{eq:sm2}
%    \end{eqnarray}
%where ${H}_I$ is the Hamiltonian of the weak interaction inducing
% the decay process. It has the form \footnote{We use
% standard notations: $\gamma^\mu$ are the Dirac matrices and
%  $\gamma^5=i\gamma^0\gamma^1\gamma^2\gamma^3$.
%  Einstein  summation over repeated indices is assumed. The metric tensor is
% $g^{\mu\nu}=\mbox{diag}(+1,-1,-1,-1)$. We refer to scalar products of
% two four-vectors $a=(a^0,\mathbf a)$  and $b=(b^0,\mathbf b)$ by
% $a{\cdot} b = a^0b^0-\mathbf{a}{\cdot}\mathbf{b}$ and use the
% Feynman slash notation $/\!\!\!a=\gamma^\mu a_\mu$.
% }
\begin{widetext}
\begin{equation}
%\mathscr
%{H}_I(\mu^-\to e^-\nu_\mu\overline\nu_{e})
S_{fi}=-i\frac{G}{\sqrt{2}}\int
[\overline{\psi}_{\nu_{\mu}}\gamma_{\lambda}(1-\gamma_{5})
\psi_{\mu}][\overline{\psi}_{e}\gamma^{\lambda}(1-\gamma_{5})\psi_{{\nu}_{e}}]d^4x.
%d^{3}{\mathbf x}.
\label{eq:sm3}
\end{equation}
\end{widetext}
Here $G=(1. 166 37\pm0. 000 02)\times{10^{-11}} MeV^{-2}$ is the
constant of the weak interaction, and $\mathbf{x}$ stands for the
spatial coordinates. $\psi_{\mu}$, $\psi_{\nu_\mu}$, $\psi_{e}$,
and $\psi_{{\nu}_{e}}$ are respectively wave functions of the
muon, the muonic neutrino, the electron, and the electronic
antineutrino.  The neutrinos are treated as massless particles
describable by  Dirac spinors \cite{Bjor64}; the minor finite-mass
effects can be included as done in Ref.\cite{Greiner}.
For the description of the laser-dressed states we note the
following. The muon, due to its large mass, is much less
influenced by the laser than the electron (for the laser
intensities considered here). The state of the electron  in the
laser field
%\textbf{(characterized by its four-momentum $p$)}
is represented by the Dirac-Volkov function (normalized in a large
volume $V$) \cite{Volkov35}. For a
 linearly polarized field it has the form
\begin{widetext}
\begin{equation}
%{\psi_q}(x)={\psi_p}(x)
\psi_{e}(x)=\left[1+\frac{e\slashed{k}\slashed{a}}{2(k{\cdot}
p)}\cos(k{\cdot} x)\right] \frac{u_e}{\sqrt{2EV}}
\exp\left[-q{\cdot} x-\frac{e(a{\cdot} p)}{k{\cdot}
p}\sin(k{\cdot} x)\right]. \label{eq:sm4}
\end{equation}
\end{widetext}
where $e$ is the electron charge, $p$ is the four-momentum of
electron for laser free, and
$q=p-\frac{e^{2}}{2(k{\cdot}p)}\overline{A^{2}}k =(E, {\bf q})$
%${{q}}^{\lambda}=p^{\lambda}-\frac{e^{2}}{2(k{\cdot}p)}\overline{A^{2}}k^{\lambda}
%=(E, {\textbf{q}})$
can be viewed as the (time) averaged four-momentum of the electron
in the presence of the laser field, with
%$e$ is the electron charge and
$\overline{A^{2}}$
%= A^2/2
%=(1/T)\int_0^T[A(x)]^2dt= -|\textbf{a}|^2/2$  being the
%=(1/T)\int_0^T[A(x)]^2dt= -{\cal E}_0^2/(2\omega^2)$
being the
%time-averaged
square of the four-potential averaged in a laser cycle.
%$T=2\pi/\omega$.
$u_e$ is a Dirac bispinor representing the free electron and it is
normalized as $\overline{u}_eu_e=2m^{2}_e$, where $\overline{u}_e$
is the Dirac adjoint of $u_e$. Inserting Eq.(\ref{eq:sm4}) and the
wave functions of mesons into Eq.(\ref{eq:sm3}) and after some
(exact) algebraic manipulations we find  the S-matrix to be
expressible as
\begin{widetext}
    \begin{eqnarray}
S_{fi}
%&=& -i\int d^{4}x~\mathscr{H}\nonumber\\
% \langle e^{-},\nu_{\mu},\overline{\nu}_{e^{-}}|:\mathscr{H}:|{\kern.5ex}\mu^{-}\rangle\kern40ex
&=&-i\displaystyle{\frac{G}{\sqrt{2}}}\sqrt{\frac{m_{\mu}m_{e}}
{E_{\mu}E}\frac{1}{2E_{\nu_{\mu}}2E_{\nu_{e}}}}\frac{1}{V^{2}}
\sum_{l}{[\overline{u}_{\nu_{\mu}}\gamma_{\lambda}(1-\gamma_{5})u_{\mu}][\overline{u}_{e}f^{\lambda}
 v_{{\nu}_e}]}
 \, {\delta(P-q-k_{\nu_{\mu}}-k_{\nu_{e}}-lk)},
    \label{eq:sm5}
    \end{eqnarray}
    \end{widetext}
where $E_{\mu}$, $E_{\nu_{\mu}}$, and $E_{\nu_{e}}$ are
respectively the energies of $\mu$, $\nu_{\mu}$ and
$\overline{\nu}_{e}$. Furthermore, $P$, $k_{\nu_{\mu}}$ and
$k_{\nu_{e}}$ are the four-momenta of $\mu$, $\nu_{\mu}$ and
$\overline{\nu}_{e}$.
%$v_{{\nu}_e}$ is the free Dirac spinor of $\overline{\nu}_{e}$.
$u_{\mu}$, $u_{\nu_{\mu}}$, and $u_{\nu_{e}}$ are respectively the
free Dirac spinors of them. In Eq.~(\ref{eq:sm5}) $l$ is the
number of photons
%
%. The explicit form of the introduced  function $f$ is
%
%\begin{widetext}
%\begin{equation}
$f^{\lambda}= (\Delta_{0}\gamma^{\lambda}+\Delta_{1}
\slashed{a}\slashed{k}\gamma^{\lambda})(1-\gamma_{5}),
\:\mbox{with}\: \Delta_{0}=J_{l}(D) \: \mbox{and}\:
%\Delta_{1}=\frac{lJ_{l}(D)}{2k{\cdot} p D},
\Delta_{1}=\frac{lJ_{l}(D)}{2(a{\cdot} p)}, \: \mbox{where}\:
D=-\frac{a{\cdot} p}{k{\cdot} p}$.
%\label{Eq:f}
%\end{equation}
%\end{widetext}
Here $J_l(D)$ is a Bessel function of order $l$.
%After the lengthily and regular calculation of QED, and
Integrating over  the electron angles ($\Omega$) and the energy
spectrum ($E$) of the final state electron we  find the decay rate
$W$
%and the lifetime $\tau_\mu$ of the muon are
is determined by the formula \footnote{We note that
Eq.(\ref{eq:sm6})
 does not account for the influence
of radiative corrections which are (in absence of the laser) small
in comparison to the modifications brought about by the laser
field.},
\begin{widetext}
%\begin {equation}
%\begin {array}{rcl}
    \begin{eqnarray}
%\frac{d{W}}{dQ_{e}}
%W
%{dW\over d\Omega}&=&
{\frac{dW}{d\Omega}}&=&
%\sum\limits_{l=-\infty}^{\infty}{dW_l\over d\Omega}
%W_l=
\sum\limits_{l=-\infty}^{\infty} \frac{G^{2}\pi}{96\pi^5}
\int_{m_e+l\omega}^{m_{\mu}/2+l\omega}dE
%\int d\Omega
\frac{|\bf{q}|}{E} \displaystyle\{\Delta_{0}^{2}[Q^{2}(P{\cdot}
p)+2(Q{\cdot} P) (Q{\cdot} p)] \: +2\Delta_{0}\Delta_{1}(a{\cdot}
p)
[Q^{2}(k{\cdot} P)\nonumber\\
&&+2(Q{\cdot} k) (Q{\cdot} P)] -2\Delta_{1}\Delta_{0}(k{\cdot} p)
[Q^{2}(a{\cdot} P) +2(Q{\cdot} a) (Q{\cdot} P)]
-2\Delta_{1}^{2}a^{2}(k{\cdot} p)[Q^{2}(k{\cdot} P)+(Q{\cdot} k)
(Q{\cdot} P)]\}.
%;\nonumber\\
%\tau_\mu &=& 1/W.
    \label{eq:sm6}
    \end{eqnarray}
%\end {array}
%\label{eq:dwdq}
%\end {equation}
\end{widetext}
Here we introduced the photon-number-resolved decay rate $W_l$ and
the momentum $Q=P-q-\emph{l\kern.1ex k}$.

\emph{Results and discussion.-} Now we discuss the numerical
result of the laser modified decay rate. The origin of the
coordinate system is chosen to be on the muon (before decay), the
z-axis is set along the direction of the electric-field vector
$\varepsilon_{0}$ of the field, and the $y$-axis is along the
direction of the wave vector $k$.

The numerous sums and integrals involved in the evaluations of
$W_l$  have to be performed numerically.
%The laser-induced bremsstrahlung and inverse
%bremsstrahlung have much great effect than the radiative
%corrections in the laser-free case.
The number of contributing multiphoton processes increases rapidly
with the field amplitude ${\cal E}_0$ or the intensity $I={\cal
E}_0^2/(8\pi)$.
\begin{figure}
%[b]
\includegraphics[height=5 cm]{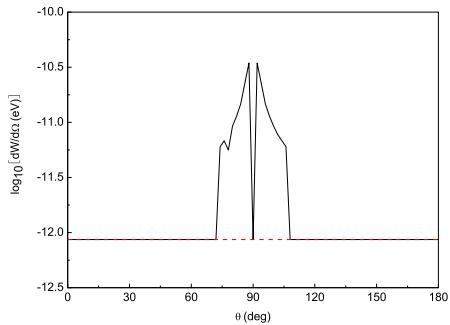}
\caption{The muon decay rate versus the polar angle at an azimuth
angle $\phi=90^\circ$
%in the rest frame of the muon
in the presence of a Nd:YAG laser ($\hbar\omega=1.17~eV$) laser
with an electric field amplitude of $10^{7}Vcm^{-1}$.}
\label{fig:1}\end{figure}
\begin{figure}
%[b]
\includegraphics[height=5 cm]{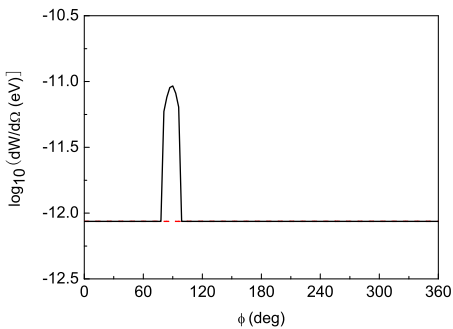}
\caption{The muon decay rate versus the azimuth angle at a polar
angle $\theta=80^\circ$. The laser parameters are the same in
Fig.1.} \label{fig:2}\end{figure}
This sets a limit on the highest $I$ we are able to consider with
our available computational resources.
%\\
%
Figs.\ref{fig:1} displays the dependence of the decay rate on the
polar angle for a Nd:YAG laser ($\hbar\omega=1.17~eV$). It is
shown that the laser modification is concentrated in a scope of
medium angle around $\theta=90^\circ$, but close
$\theta=90^\circ$, the laser effect vanishes soon. At other
angles, the laser effect vanishes. This suggest that the
intermediate meson muon may have internal structure. If the meson
has completely no structure as the standard model suggested, the
laser-free decay should be isotropic; When the laser is presented,
the decay should be enlarged in the polarization direction.

In Fig.~\ref{fig:2} we  show the decay rate versus the azimuth angle.
The decay rate is enlarged around $\theta=90^\circ$, i.e. along
the direction of the photon momentum of the laser. This suggest
the photons are directly interact with the components in the muon,
and affect the angular distribution of final state electron.

In summary we provided theoretical evidence that the decay of muon
can be modified by a linearly polarized laser field. It suggests
muon may have structures, and indicates an approach to study the
structure by applying a strong laser background. This effect
deserves further investigation both experimentally and
theoretically.

 {\it Acknowledgments.-}
  This work was supported by the National Natural Science Foundation
of China under Grant Numbers 10674125 and 10475070.

\clearpage
%  \section*{FIGURE CAPTIONS}
%


\begin{thebibliography}{s2}
    %
      \bibitem{sj0} G. A. Mourou, C. P. J. Barty, and M. D. Perry, Phys. Today
\textbf{51}, 22 (1998);  S.V. Bulanov \emph{et al.},
% T. Esirkepov, and T. Tajima,
Phys. Rev. Lett. \textbf{91}, 085001 (2003); S. Gordienko \emph{et
al.},
%,. Pukhov, O. Shorokhov, and T. Baeva,
 Phys. Rev. Lett. \textbf{94}, 103903 (2005).
 %; J. Nees et al., J. Mod. Opt. 52, 305 (2005).
    %
     \bibitem{sj3} R. Snavely \emph{et al.}, Phys. Rev. Lett. \textbf{85}, 2945
     (2000) and references therein.
     %
         %
    \bibitem{sj1} Toma Toncian \emph{et al.}, Science \textbf{312}, 410
    (2006);  B. M. Hegelich \emph{et al.}, Nature \textbf{439}, 441
    (2006);  H. Schwoerer \emph{et al.}, Nature \textbf{439}, 445 (2006).
%
     %
        \bibitem{sj7} A. Rousse \emph{et al.}, Phys. Rev. Lett. \textbf{93}, 135005 (2004).
        %
    \bibitem{sj8} K. W. D. Ledingham, P. McKenna, R. P. Shinghal, Science \textbf{300}, 1107
    (2003); D. Umstadter, Nature (London) \textbf{404}, 239 (2000).

    \bibitem{sj9} B. A. Remington, D. Arnet, R. P. Drake, H. Takabe, Science \textbf{284},
1488 (1999); G. A. Mourou, T. Tajima, and S.V. Bulanov, Rev. Mod.
Phys. \textbf{78}, 309 (2006); M. Marklund and P. K. Shukla,
\emph{ibid}. \textbf{78}, 591 (2006).
%
%
\bibitem{Chelkowski} S. Chelkowski, A. D. Bandrauk, and P. B. Corkum, Phys. Rev. Lett.
\textbf{93}, 083602 (2004),
%
\bibitem{Griffiths}David J.  Griffiths, \emph{Introduction to Elementary Particles}
%(Wiley, John \& Sons, Inc., 1987)
(John Wiley \& Sons,  New York, 1987)
%
\bibitem{Brom} D. A. Bromley,  \emph{Gauge Theory of Weak Interactions}
(Springer, Berlin, 2000).
%
  %  \bibitem{Cite02} I.   I.   Goldman,  Phys.   Lett.   {\bf 8}, 103 (1964).
%    \bibitem{Cite021}A. I. Goldman, Zh.  \'{E}ksp.  Teor.  Fiz.  {\bf 46}.  1412 (1964) [Sov.  Phys.  JETP {\bf 19},
%    954 (1964)].
%    \bibitem{Cite022} A.   I.   Nikishov and V.   I.   Ritus, Zh. \'Eksp. Teor. Fiz.  {\bf 46},  1768 (1964) [Sov.  Phys.  JETP {\bf
%    19}, 119].
%    \bibitem{Cite023} N.   B.   Narozhny,  A.   I.   Nikishov, and V.   I.   Ritus, Zh. \'Eksp.  Teor.  Fiz.  {\bf 47}, 930 (1964) [Sov.  Phys.  JETP  {\bf 20}, 622 (1965)].  % more
%    \bibitem{Cite01} L.   S. Brown and T.   W.   B.   Kibble, Phys.   Rev.  A {\bf
%    133},  705  (1965).
%    \bibitem{Cite011} T.   W.   B.   Kibble,  Phys.   Rev.   {\bf 150},  1060
%     (1966).
%     \bibitem{Cite012} J.  H.  Eberly and H.  R.  Reiss,  Phys.  Rev. {\bf 145},  1035  (1966).
%     \bibitem{Cite013} H.  R.  Reiss and J.  H.  Eberly,  Phys.  Rev.  {\bf 151},  1058  (1966).
%     \bibitem{Cite014} J. H.  Eberly and A.  Sleeper,  Phys.  Rev.  {\bf 176},  1570  (1968).
%     \bibitem{Mitt93} M.   H.   Mittleman,  {\it Introduction to the Theory of Laser-Atom Interactions}  (Plenum,  New York,  1993).
%
%    \bibitem{Fran90} P.   Francken and C.   J.   Joachain,  J.   Opt.   Soc.   Am.   B {\bf 7},  554  (1990).
%    \bibitem{Ehlo98} F.   Ehlotzky,  A.   Jaro\'{n},  and J.   Z.   Kami\'{n}ski,  Phys.   Rep.   {\bf 297},  63  (1998).
%     \bibitem{Ehlo01} F.   Ehlotzky,  Phys.   Rep.   {\bf 345},  175  (2001).
%    %
%    \bibitem{Ehlo88}F.   Ehlotzky,  Opt.   Commun.   {\bf 66},  265  (1988).
%    \bibitem{Kami99}J.   Z.   Kamin\'{s}ki and F.   Ehlotzky,  Phys.   Rev.   A {\bf 59},  2105  (1999).
%    \bibitem{Pane99}P.   Panek,  J.  Z.   Kamin\'{s}ki,  and F.   Ehlotzky,  Can.   J.   Phys.   {\bf 77},  591
% (1999).
%    \bibitem{Deni} M.  M.   Denisov and M.   V.   Fedorov,  Zh.   {\'E}ksp.   Teor.   Fiz.   {\bf 53},
%1340 (1967) [Sov.   Phys.   JETP {\bf 26},  779 (1968)].
%    \bibitem{Rosh} S.  P.   Roshchupkin,  Laser Phys.   {\bf 3},  414 (1993); Laser Phys. {\bf 7},  873 (1997);
%Zh.   Eksp.   Teor.   Fiz.   {\bf 106},  102 (1994) [Sov.   Phys.
%JETP {\bf 79}, 54 (1994)]; Zh.   Eksp.   Teor.   Fiz.  {\bf 109},
%337 (1996) [Sov. Phys. JETP {\bf 82},  177 (1996)].
%%
%%    \bibitem{Szym97} C.   Syzmannowski,  V.   V\'{e}niard, R.  Ta\"{\i}eb,  and
%%    A.  Maquet,  C.  H.  Keitel,  Phys.  Rev.  A.  {\bf 56}, 3846 (1997).
%    \bibitem{Szym97} C.   Szymanowski,  V.   V\'{e}niard,  R.   Ta{\"{\i}}eb,  A.   Maquet,
%and C.   H.   Keitel,  Phys.   Rev.   A {\bf 56},  3846  (1997).
%    \bibitem{Szym98} C.   Szymanowski and A.   Maquet,  Opt.   Express {\bf 2},  262  (1998).
%\bibitem{Lism03} S.-M.   Li,  J.   Berakdar,  J.   Chen,  and Z.-F.   Zhou,  Phys.   Rev.   A {\bf 67},  063409  (2003).
%    \bibitem{Atta} Y.   Attaourti and B.   Manaut,  Phys.   Rev.   A {\bf 68},  067401
%    (2003).
%    \bibitem{Atta01}Y.   Attaourti,  B.   Manaut,  and A.   Makhoute,  Phys.   Rev.   A   {\bf 69},  063407
%    (2004).
%    \bibitem{Atta02}Y.   Attaourti and S.   Taj,  Phys.   Rev.   A  {\bf 69},  063411
%    (2004).
 %   \bibitem{Atta03}Y.  Attaourti,  B.   Manaut,  and S.   Taj,  Phys.   Rev.   A  {\bf 70},  023404  (2004).
 %
    \bibitem{PDG06} S.  Eidelman \emph{et al.},
    %(Particle Data Group,  URL: http://pdg. lbl. gov ).  Review of Particle Physics.
    Phys.   Lett.   B  {\bf 592},  1  (2004).

    \bibitem{Greiner} W.  Greiner and  B.  M{\"{u}}ller,  {\it Gauge Theory of
    the Weak Interaction}  (Springer,  Berlin,  2003).

    \bibitem{Czar00} A.   Czarnecki,  G.   P.   Lepage, and  W.   J.   Maraciano,
    Phys.   Rev.   D {\bf 61},  073001 (2000).

%     \bibitem{Vshi79} A.  S.  Vshivtsev,  P.  A.  \'{E}minov, Theo.  Math.  Phys,  Vol44,  Num 2, pp284-288, Aug,   (1980).
     \bibitem{Vshi79} A.  S.  Vshivtsev and  P.  A.  \'{E}minov, Theo.  Math.  Phys {\bf 44} (2), 284 (1980).




%    \bibitem{Chu88} S.   Chu,  A.  P.  Mills Jr.,  A.   G.   Yodh,  K.   Nagamine,  Y.   Miyake,
%    and T.   Kuga,  Phys.   Rev.   Lett {\bf 60},  101 (1988).
    \bibitem{Chu88} S.   Chu  \emph{et al.},  Phys.   Rev.   Lett {\bf 60},  101 (1988).

%    \bibitem{Naga95} K.   Nagamine,  Y.   Miyake,  K.   Shimomura,  P.   Birrer,  J.   P.   Marangos,
%    M.   Iwasaki,  P.   Strasser,  and T.  Kuga,  Phys.   Rev.   Lett {\bf 74},  4811 (1995).
    \bibitem{Naga95} K.   Nagamine \emph{et al.},  Phys.   Rev.   Lett {\bf 74},  4811 (1995).

     \bibitem{Milc96}H.  M.   Milchberg and R.  R.   Freeman,  J.   Opt.   Soc.   Am.   B {\bf 13},  51  (1996).

%    \bibitem{PDG06} S.  Eidelman $et~al$,  (Particle Data Group,  URL: http://pdg. lbl. gov ).  Review of Particle Physics.
%    Phys.   Lett.   B  {\bf 592},  1  (2004).

%    \bibitem{rep_pro_phys} J.   Ullrich {\it et al.  },  Reports on Progress
%in Physics 66.   1463  (2003).

%    \bibitem{Bula96} C. Bula, K. T. McDonald, and E. J. Prebys, C. Bamber, S. Boege,
%    T. Kotseroglou, A. C. Melissinos, D. D. Meyerhofer, and W. Ragg, D. L. Burke,
%    R. C. Field, G. Horton-Smith, A. C. Odian, J. E. Spencer, and D. Walz, S. C. Berridge,
%    W. M. Bugg, K. Shmakov, and A. W. Weidemann,  Phys.   Rev.   Lett.   {\bf 76},  3116  (1996).
    \bibitem{Bula96} C. Bula \emph{et al.},  Phys.   Rev.   Lett.   {\bf 76},  3116  (1996).

    \bibitem{Bjor64} J.   D.  Bjorken and  S.   D.   Drell,  {\it Relativistic Quantum
    Mechanics}  (McGraw-Hill,  New York,  1964).

%    \bibitem{iwaa06} Aihua Liu,  Shumin Li,  e-print physics/0000000

    \bibitem{Volkov35} D.   M.   Volkov,  Z.   Phys.   {\bf 94},  250  (1935).


%    \bibitem{Szy97} C. Szymanowski, V. V\'{e}niard, R. Ta{\"{\i}}eb, A. Maquet,
%and C. H. Keitel, Phys. Rev. A {\bf 56}, 3846 (1997).
    \bibitem{Szy97} C. Szymanowski \emph{et al.}, Phys. Rev. A {\bf 56}, 3846 (1997).


\bibitem{Lism03} S.-M. Li, J. Berakdar, J. Chen, and Z.-F. Zhou, Phys. Rev. A {\bf 67}, 063409 (2003).

%
    \bibitem{Fedo97} M.   V.   Fedorov,  {\it Atomic and Free Electrons in a Strong Light
    Field} (World Scientific,  Singapore, 1997).

   \bibitem{Keldysh}
   L. V. Keldysh,
   %''Ionization in the field of a strong
%electromagnetic wave,''
% Sov. Phys. JETP \textbf{20}, 1307?314 (1965).
Sov. Phys. JETP \textbf{20}, 1307 (1965).

   \bibitem{prel}M. Perelomov, V. S. Popov, and M. V. Terent'ev,
 % ''Ionization of atoms
% in an alternating electric field,''
%Sov. Phys. JETP \textbf{23}, 924?34 (1966).
Sov. Phys. JETP \textbf{23}, 924 (1966).
    \end{thebibliography}
\end{document}